\documentclass[a4paper,11pt,fleqn]{article}

\usepackage[ansinew]{inputenc}
\usepackage[mathscr]{eucal}
\usepackage{amsmath,amssymb,amsthm}
\usepackage{graphicx}
\usepackage{color}

\allowdisplaybreaks

\makeatletter
\long\def\@makecaption#1#2{%
  \vskip\abovecaptionskip\footnotesize
  \sbox\@tempboxa{#1. #2}%
  \ifdim \wd\@tempboxa >\hsize
    #1. #2\par
  \else
    \global \@minipagefalse
    \hb@xt@\hsize{\hfil\box\@tempboxa\hfil}%
  \fi
  \vskip\belowcaptionskip} 
\makeatother

\newcommand{\dd}[2]{\frac{\mathrm{d} #1}{\mathrm{d} #2}}

\newcommand{\ddd}{\mathrm{d}}

\newcommand{\p}{\partial}

\newcommand{\todo}[1][\null]{\ensuremath{\clubsuit}}

\newcommand{\noprint}[1]{}

{\theoremstyle{definition}

\newtheorem*{remark*}{Remark}
}

\newcommand{\lsemioplus}{\mathbin{\mbox{$\lefteqn{\hspace{.77ex}\rule{.4pt}{1.2ex}}{\in}$}}}

\newcommand{\checked}[1][\null]{\ensuremath{\boldsymbol{\surd}}}

\newcommand{\DD}{\mathrm{D}}

\newcommand{\ve}{\varepsilon}

\newcommand{\Ad}{\mathrm{Ad}}

\newcommand{\ZZ}{\mathcal{Z}}

\newcommand{\DDD}{\mathcal{D}}
\newcommand{\JJJ}{\mathcal{J}}

\begin{document}

\par\noindent {\LARGE\bf
Lie reduction and exact solutions of\\ vorticity equation on rotating sphere
\par}

{\vspace{4mm}\par\noindent {\bf Alexander Bihlo$^\dag$ and Roman O.\ Popovych$^\ddag$
} \par\vspace{2mm}\par}

{\vspace{2mm}\par\noindent {\it
$^{\dag}$~Centre de recherches math\'{e}matiques, Universit\'{e} de Montr\'{e}al, C.P.\ 6128, succ.\ Centre-ville,\\
$\phantom{^\dag}$~Montr\'{e}al (QC) H3C 3J7, Canada
}}

{\vspace{2mm}\par\noindent {\it
$^{\ddag}$~Wolfgang Pauli Institute, Nordbergstra{\ss}e 15, A-1090 Vienna, Austria\\
$\phantom{^\ddag}$~Institute of Mathematics of NAS of Ukraine, 3 Tereshchenkivska Str., 01601 Kyiv, Ukraine
}\par}

{\vspace{2mm}\par\noindent {\it
$\phantom{^\dag}$~\textup{E-mail}:
$^{\dag}$bihlo@crm.umontreal.ca,
$^{\ddag}$rop@imath.kiev.ua
}\par}

\vspace{4mm}\par\noindent\hspace*{5mm}\parbox{150mm}{\small
Following our paper [J.\ Math.\ Phys.\ {\bf 50} (2009) 123102], we systematically carry out Lie symmetry analysis for the barotropic vorticity equation on the rotating sphere.
All finite-dimensional subalgebras of the corresponding maximal Lie invariance algebra, which is infinite-dimensional, are classified.
Appropriate subalgebras are then used to exhaustively determine Lie reductions of the equation under consideration.
The relevance of the constructed exact solutions for the description of real-world physical processes is discussed.
It is shown that the results of the above paper are directly related to the results of the recent letter by N.~H.~Ibragimov and R.~N.~Ibragimov [Phys.\ Lett.\ A {\bf 375} (2011) 3858]
in which Lie symmetries and some exact solutions of the nonlinear Euler equations for an atmospheric layer in spherical geometry were determined.
}\par\vspace{2mm}

\section{Introduction}

One approach to reduce the complexity of the governing equations of geophysical fluid dynamics is to consider submodels thereof. The latter usually take into account several simplifications that make them adapted for distinct physical processes or that render them valid only on well-defined spatial or temporal scales. One of the most relevant simplifications which can be reasonably implemented in the atmospheric sciences is that of the two-dimensionality of large-scale flow. This reduction in dimensionality greatly simplifies the structure of the dynamical part of the governing equations, which can be consistently described by some forms of the two-dimensional Euler or Navier--Stokes equations~\cite{andr98Ay,fush94Ay,ibra95Ay,mele05Ay,olve86Ay,pukh06Ay}. Though they are still coupled systems of nonlinear partial differential equations, much is known by today about the long-term behavior of these systems.

A form of the two-dimensional Euler or Navier--Stokes equations which is of immediate interest in large-scale atmospheric dynamics is that in spherical geometry. In the recent letter~\cite{ibra11Ay} the system of nonlinear incompressible Euler equations in a single thin atmospheric layer on the sphere rotating with constant angular velocity~$\Omega$ was considered,
\begin{align}\label{eq:NonlinearEulerEquationsSphere}
\begin{split}
 &u_t+\frac{u}{\sin\theta}\,u_\lambda+vu_\theta+v u\cot\theta +\frac{v\cos\theta}{R_0}+\frac{1}{\sin\theta}\,p_\lambda=0,\\
 &v_t+\frac{u}{\sin\theta}\,v_\lambda+vv_\theta-u^2\cot\theta -\frac{u\cos\theta}{R_0}+p_\theta=0,\\
 &u_\lambda+(v\sin\theta)_\theta=0.
\end{split}
\end{align}
In the above system $\lambda$ and $\theta$ denote the azimuthal and the polar angles, respectively. The former increases by going East, while the latter increases by going South. The associated velocity components are $u$ and $v$, respectively, $p$ denotes the pressure and $R_0$ is the Rossby number, which is the ratio of inertial to Coriolis force, $R_0\propto \Omega^{-1}$. The last equation of the above system is the incompressibility condition. Subscripts as usual denote the partial derivatives.

In~\cite{ibra11Ay} it was assumed that
\begin{equation}\label{eq:AnsatzStreamfunctionIbragimov}
 u=F(\theta) + \nu\hat u,\quad v = \nu\hat v,\quad p=\bar p + \nu\hat p,
\end{equation}
where $\nu$ is a (not necessarily small) parameter, $F(\theta)$ is an arbitrary function of its argument and $\bar p$ is the mean pressure in the thin atmospheric layer studied. Variables with hats are to be interpreted as disturbance quantities. Physically, this ansatz corresponds to finite disturbances that are superimposed to a zonally averaged mean flow having only a latitudinal dependence. By introducing a stream function $\hat \psi$ for the disturbance velocities, i.e.\
\[
 \hat u=\hat\psi_\theta,\qquad \hat v = - \frac{1}{\sin\theta}\,\hat\psi_\lambda,
\]
and taking the curl of the first two equations of the system~\eqref{eq:NonlinearEulerEquationsSphere}, one arrives at
\begin{align}\label{eq:VorticityEquationIbragimov}
\hat\zeta_t + \frac{\nu}{\sin\theta}\,(\hat\psi_\theta\hat\zeta_\lambda-\hat\psi_\lambda\hat\zeta_\theta)+\frac{F}{\sin\theta}\,\hat\zeta_\lambda - \frac{1}{\sin\theta}\,\hat\psi_\lambda L_1F+\frac{1}{R_0}\,\hat\psi_\lambda = 0,
\end{align}
where the vorticity~$\hat \zeta$ is related to the stream function~$\hat \psi$ through the Laplacian in spherical geometry,
\[
\hat\zeta:=\frac{1}{\sin\theta}\,(\sin\theta\hat\psi_\theta)_\theta+\frac{1}{\sin^2\theta}\,\hat\psi_{\lambda\lambda},
\]
and $L_1$ denotes the Sturm--Liouville operator for the associated Legendre functions,
\[
 L_1:=\frac{1}{\sin\theta}\dd{}{\theta}\left(\sin\theta\dd{}{\theta}\right)-\frac{1}{\sin^2\theta}.
\]
This is the equation considered in~\cite{ibra11Ay}.

Previously, in~\cite{bihl09Ay} we have studied the Lie symmetries and exact solutions of the well-known barotropic vorticity equation on the rotating sphere,
\begin{equation}\label{eq:VorticityEquationSphere}
 \zeta_t + \psi_\lambda\zeta_\mu - \psi_\mu\zeta_\lambda + 2\Omega\psi_\lambda = 0,\qquad
 \zeta := \frac{1}{1-\mu^2}\,\psi_{\lambda\lambda}+((1-\mu^2)\psi_\mu)_\mu,
\end{equation}
which is also derived from the system of nonlinear Euler equations~\eqref{eq:NonlinearEulerEquationsSphere}, but without introducing the ansatz~\eqref{eq:AnsatzStreamfunctionIbragimov}.
In the above equation the variable~$\mu$ is defined as $\mu=\sin\varphi$ with $\varphi$ being the geographic latitude on the sphere. The mean radius of the Earth is assumed to be scaled to the unity.

It is the purpose of the present letter to relate the above two independent studies. In Section~\ref{sec:RelationIbragimovBarotropicVorticity} we explicitly transform Eqs.~\eqref{eq:VorticityEquationIbragimov} and~\eqref{eq:VorticityEquationSphere} to each other. This allows us to derive the results on the Lie symmetries and reductions of Eq.~\eqref{eq:VorticityEquationIbragimov} obtained in~\cite{ibra11Ay} from our previous work~\cite{bihl09Ay}. The related Lie reductions and exact solutions of Eq.~\eqref{eq:VorticityEquationSphere} are discussed in Section~\ref{sec:RelationAlgebrasIbragimovBarotropicVorticity}. Additionally, in Section~\ref{sec:Subalgebras} we present the exhaustive classification of finite-dimensional subalgebras of the maximal Lie invariance algebra of Eq.~\eqref{eq:VorticityEquationSphere}. In Section~\ref{sec:Conclusion} we give some concluding general remarks on Lie reduction and illustrate them with the reductions carried out for the nonlinear Euler equations in a single thin atmospheric layer.

\section{Relation between the equations}\label{sec:RelationIbragimovBarotropicVorticity}

It is straightforward to show that Eq.~\eqref{eq:VorticityEquationIbragimov} and Eq.~\eqref{eq:VorticityEquationSphere} can be mapped to each other by point transformations. In fact, their difference merely comes from a different set of variables employed and the special ansatz~\eqref{eq:AnsatzStreamfunctionIbragimov} adopted in~\cite{ibra11Ay}.

First we should like to note that in geophysical fluid dynamics it is common to use the latitude $\varphi$ rather than the polar angle $\theta$ as an independent variable. They are related by $\varphi=\pi/2-\theta$, hence $\mu=\sin\varphi=\cos\theta$. The second difference between the two forms of the vorticity equation stems from the ansatz~\eqref{eq:AnsatzStreamfunctionIbragimov} assumed in~\cite{ibra11Ay}. This ansatz plays no essential role for the study of nonlinear Euler equations as it is explicitly assumed that $\nu$ is not a small parameter. Therefore it is possible to avoid the separation of~$F(\theta)$ from the disturbance velocity $\hat v$ and the introduction of the parameter expansion in~$\nu$. As shown in Section~\ref{sec:RelationAlgebrasIbragimovBarotropicVorticity}, this ansatz even complicates the structure of exact solutions that can be derived. Solutions in the form of a separated mean flow superimposed by some disturbance can be generated by the exact solutions that can be derived from the usual form of Eq.~\eqref{eq:VorticityEquationSphere}.
In fact, setting
\begin{equation}\label{eq:RelationStreamFunctions}
 \hat \psi = \frac{1}{\nu}\left(\psi-\int F\ddd\theta\right)
\end{equation}
and using the triple of independent variables $(t,\lambda,\mu)$ rather than $(t,\lambda,\theta)$ leads to the cancelation of the terms involving~$F$ and thus to the classical barotropic vorticity equation in a rotating reference frame~\eqref{eq:VorticityEquationSphere}. Note that the above transformation is singular for $\nu=0$, a case which yields an exact solution of the incompressible Euler equations for $F=1/\sin\theta$ as discussed in~\cite{ibra11Ay}.

A further simplification of Eq.~\eqref{eq:VorticityEquationSphere} can be achieved upon using the transformation
\begin{equation}\label{eq:PlatzmanTransformation}
 \tilde t=t,\quad \tilde \mu=\mu,\quad \tilde \lambda = \lambda+\Omega t,\quad \tilde \psi = \psi-\Omega\mu,
\end{equation}
which was first found in~\cite{plat60Ay} and then rediscovered in~\cite{bihl09Ay} using methods of group analysis. This transformation allows mapping the vorticity equation in a rotating reference frame to the corresponding equation in a reference frame at rest, i.e.\ to set $\Omega=0$ in Eq.~\eqref{eq:VorticityEquationSphere}.

\section{Subalgebras of maximal Lie invariance algebra}\label{sec:Subalgebras}

Taking into account all the transformations discussed in Section~\ref{sec:RelationIbragimovBarotropicVorticity}, the maximal Lie invariance algebra presented in~\cite{ibra11Ay} is of course isomorphic to the maximal Lie invariance algebra~$\mathcal S^\infty_0$ of the vorticity equation~\eqref{eq:VorticityEquationSphere} for $\Omega=0$, which is generated by the vector fields~\cite{bihl09Ay}
\begin{align}\label{eq:MaximalLieInvarianceAlgebraVorticityEquationSphere}
\begin{split}
 &\DDD=t\p_t-\psi\p_\psi,\quad\p_t,\quad \JJJ_1=\p_\lambda,\quad \JJJ_2=\mu\frac{\sin\lambda}{\sqrt{1-\mu^2}}\p_\lambda+\sqrt{1-\mu^2}\cos\lambda\p_\mu,\\
 &\JJJ_3=\mu\frac{\cos\lambda}{\sqrt{1-\mu^2}}\p_\lambda-\sqrt{1-\mu^2}\sin\lambda\p_\mu,\quad \ZZ(g)=g(t)\p_\psi,
\end{split}
\end{align}
where the parameter~$g$ traverses the set of smooth functions of~$t$. The algebra~$\mathcal S^\infty_0$ has the structure of $\mathfrak{so}(3)\oplus(\mathfrak g_2\lsemioplus\langle\ZZ(g)\rangle)$, i.e.\ it is the direct sum of $\mathfrak{so}(3)$ (with the basis elements~$\JJJ_1$, $\JJJ_2$ and~$\JJJ_3$) and the semidirect sum of the two-dimensional non-Abelian algebra $\mathfrak g_2=\langle\DDD,\p_t\rangle$ with the infinite-dimensional Abelian ideal $\langle\ZZ(g)\rangle$.
Eq.~\eqref{eq:VorticityEquationSphere} admits also two independent discrete symmetries, $(t,\lambda,\mu,\psi)\mapsto(-t,-\lambda,\mu,\psi)$ and $(t,\lambda,\mu,\psi)\mapsto(t,\lambda,-\mu,-\psi)$.

In order to systematically carry out Lie reduction, an optimal list of low-dimensional subalgebras of the maximal Lie invariance algebra~$\mathcal S^\infty_0$ is required, which was also determined in~\cite{bihl09Ay}.
As subalgebras of greater dimensions may be useful, e.g.\ in the study of partially and differentially invariant solutions~\cite{andr98Ay,golo04Ay,mele05Ay,ovsi82Ay} and invariant parameterizations~\cite{bihl11Fy,ober97Ay,popo10Cy,raza07By},
here we classify all finite-dimensional subalgebras of~$\mathcal S^\infty_0$.
Up to equivalence generated by the internal automorphisms of~$\mathcal S^\infty_0$, a complete list of such subalgebras is exhausted by the following parameterized classes of subalgebras:
\begin{gather*}
1)\, I^n(\bar g),\quad
2)\, \langle\JJJ_1{+}\ZZ(f)\rangle\oplus I^n(\bar g),\quad
3)\, \langle\JJJ_1,\,\JJJ_2,\,\JJJ_3\rangle\oplus I^n(\bar g),
\\
4)\, \langle\p_t{+}\sigma\JJJ_1\rangle\oplus\hat I^{nn'}_{\bar\lambda\bar m\tilde\mu\tilde\nu\tilde m'},\
5)\, \langle\p_t,\,\JJJ_1{+}\ZZ(\kappa t^k)\rangle\oplus\hat I^{nn'}_{\bar\lambda\bar m\tilde\mu\tilde\nu\tilde m'},\
6)\, \langle\p_t,\,\JJJ_1,\,\JJJ_2,\,\JJJ_3\rangle\oplus\hat I^{nn'}_{\bar\lambda\bar m\tilde\mu\tilde\nu\tilde m'},
\\
7)\, \langle\DDD{+}\sigma\JJJ_1\rangle\oplus\check I^{nn'}_{\bar\lambda\bar m\tilde\mu\tilde\nu\tilde m'},\
8)\, \langle\DDD,\,\JJJ_1{+}\ZZ(\kappa t^{-1}\tau^k)\rangle\oplus\check I^{nn'}_{\bar\lambda\bar m\tilde\mu\tilde\nu\tilde m'},\
9)\, \langle\DDD,\,\JJJ_1,\,\JJJ_2,\,\JJJ_3\rangle\oplus\check I^{nn'}_{\bar\lambda\bar m\tilde\mu\tilde\nu\tilde m'},
\\
10)\, \langle\DDD{+}\sigma\JJJ_1{+}\ZZ(\kappa t^n),\,\p_t\rangle\oplus\tilde I^n\!,\
11)\, \langle\DDD,\,\p_t,\,\JJJ_1{+}\ZZ(\kappa t^n)\rangle\oplus\tilde I^n\!,\
12)\, \langle\DDD,\,\p_t,\,\JJJ_1,\,\JJJ_2,\,\JJJ_3\rangle\oplus\tilde I^n\!,
\end{gather*}
where for convenience we use several notations for subalgebras contained in the ideal~$\langle\ZZ(g)\rangle$,
\begin{gather*}
I^n(\bar g)=\langle\ZZ(g^1),\,\dots,\,\ZZ(g^n)\rangle,\\
\hat I^{nn'}_{\bar\lambda\bar m\tilde\mu\tilde\nu\tilde m'}= \langle\ZZ(t^{k_i}e^{\lambda_it}),\,
\ZZ(t^{l_j}e^{\mu_jt}\cos\nu_jt),\,\ZZ(t^{l_j}e^{\mu_jt}\sin\nu_jt),\, ^{k_i=0,\dots,m_i}_{{\smash l}_j=0,\dots,m'_{\smash j}},\,{}^{i=1,\dots,n}_{j=1,\dots,n'}\rangle,\\
\check I^{nn'}_{\bar\lambda\bar m\tilde\mu\tilde\nu\tilde m'}= \langle\ZZ(\tau^{k_i}|t|^{\lambda_i}),\,
\ZZ(\tau^{l_j}|t|^{\mu_j}\cos(\nu_j\tau)),\,\ZZ(\tau^{l_j}|t|^{\mu_j}\sin(\nu_j\tau)),\, ^{k_i=0,\dots,m_i}_{{\smash l}_j=0,\dots,m'_{\smash j}},\,{}^{i=1,\dots,n}_{j=1,\dots,n'}\rangle,\\
\tilde I^n=\langle\ZZ(1),\,\dots,\,\ZZ(t^{n-1})\rangle.
\end{gather*}
Here $\bar g=(g^1,\dots, g^n)$ is an arbitrary $n$-tuple of linearly independent functions of~$t$,
$n,n'\in\mathbb N_0$, $\bar m\in\mathbb N_0^n$, $\tilde m'\in\mathbb N_0^{n'}$,
$\bar\lambda$ is an $n$-tuple of different real numbers,
$\tilde\mu$ and~$\tilde\nu$ are $n'$-tuples of real numbers such that all pairs $(\mu^j,\nu^j)$ are different,
$\tau=\ln|t|$.
In the fifth (resp.\ eighth) class we have $k=k_i$ with the value of~$i$ such that $\lambda_i=0$ ($\lambda_i=-1$) and $k=0$ if such a value of~$i$ does not exist.
Re-combining the basis elements, we also can assume that
the function~$f$ is either identically equal to zero or linearly independent with the functions~$g^1$, \dots, $g^n$.
There are equivalences between different subalgebras within the above classes induced by shifts and scalings of~$t$.
Thus, the argument~$t$ can be replaced in~$\bar g$  and~$f$ by $e^{\varepsilon_1}t+\varepsilon_0$, where $\varepsilon_0,\varepsilon_1\in\mathbb R$.
In classes 4--6, a single nonzero parameter among $\sigma$, $\lambda_i$, $\mu_j$, $\nu_j$ and $\kappa$ can be scaled to~$\pm1$.
We also can set $\kappa=\pm1$ for classes~10 and~11 if $\kappa\ne0$.
An additional possibility for constraining signs of parameters is given by discrete symmetries of Eq.~\eqref{eq:VorticityEquationSphere}.

We omit the proof of the classification result because it is too cumbersome.

\section{Lie reduction and exact solutions}\label{sec:RelationAlgebrasIbragimovBarotropicVorticity}

As the vorticity equation~\eqref{eq:VorticityEquationSphere} is an equation in three independent variables, Lie reduction using one- and two-dimensional subalgebras will lead to submodels of Eq.~\eqref{eq:VorticityEquationSphere} that are partial differential equations in two independent variables and ordinary differential equations, respectively.
Recall that it suffices to study only Eq.~\eqref{eq:VorticityEquationSphere} with $\Omega=0$.

First we select inequivalent one-dimensional subalgebras of~$\mathcal S^\infty_0$, which are appropriate for Lie reduction.
An optimal list of one-dimensional subalgebras of~$\mathcal S^\infty_0$ reads
\begin{align*}
    \langle \DDD +a\JJJ_1\rangle,\qquad \langle \p_t+a\JJJ_1\rangle, \qquad \langle\JJJ_1 + \ZZ(g) \rangle,\qquad \langle \ZZ(g) \rangle,
\end{align*}
where $a\in \mathbb{R}$ and $a\in \{-1,0,1\}$ for the first and second cases, respectively,
and $g$ is an arbitrary smooth function of~$t$, which does not vanish in the last case.
See~\cite{bihl09Ay} for more details.
The algebra $\langle \ZZ(g) \rangle$ cannot be used for classical Lie reduction as it does not permit making a reduction ansatz for~$\psi$.
This is why we only have to consider the first three subalgebras.

For the subalgebra $\langle\DDD + a\JJJ_1 \rangle$ the invariants of the associated one-parametric group are $p = \lambda -a\ln t$, $q = \mu$ and $v=t\psi$.
The reduction ansatz therefore is $\psi=t^{-1}v(p,q)$, which yields
\begin{equation}\label{eq:VorticityEquationSphereFirstReduction}
w + aw_p - v_pw_q + v_qw_p = 0, \quad w := \frac1{1-q^2}v_{pp} + ((1-q^2)v_q)_q
\end{equation}
as a submodel of~\eqref{eq:VorticityEquationSphere}. Though we have reduced the number of independent variables by one, it appears to be impossible to completely solve Eq.~\eqref{eq:VorticityEquationSphereFirstReduction}. Solutions of Eq.~\eqref{eq:VorticityEquationSphereFirstReduction} can be sought, for example, using numerical integration. We will return to a further reduction of Eq.~\eqref{eq:VorticityEquationSphereFirstReduction} when we discuss reductions with respect to two-dimensional subalgebras.

An ansatz constructed using the subalgebra $\langle \p_t+a\JJJ_1\rangle$ is $\psi = v(p,q)$, where $p = \lambda - at$ and $q = \mu$. This ansatz reduces Eq.~\eqref{eq:VorticityEquationSphere} to
\begin{equation}\label{eq:RossbyHaurwitzClass}
 -(v+aq)_qw_p + (v+aq)_pw_q = 0,\quad w := \frac1{1-q^2}v_{pp} + ((1-q^2)v_q)_q.
\end{equation}
Eq.~\eqref{eq:RossbyHaurwitzClass} implies that $w=F(v+aq)$, for $F$ being an arbitrary smooth function of its argument. The definition of~$w$ then takes the form of a nonlinear Poisson equation.
In the particular case of $F$ being a linear function, i.e.\ $F=c(v+aq)+b$, where $b$ and $c$ are real constants, this equation is a linear inhomogeneous PDE,
\begin{equation}\label{eq:RossbyHaurwitzLinearEq}
\frac1{1-q^2}v_{pp} + ((1-q^2)v_q)_q=cv+caq+b.
\end{equation}
The general solution of Eq.~\eqref{eq:RossbyHaurwitzLinearEq} is represented in the form $v=v_{\rm g.h.}+v_{\rm p.i.}$,
where $v_{\rm p.i.}$ is a particular solution of Eq.~\eqref{eq:RossbyHaurwitzLinearEq}
and $v_{\rm g.h.}$ is the general solution of the corresponding homogenous equation, which we denote by~$\mathcal L$.
As $v_{\rm p.i.}$ we can take the following solutions:
\begin{gather*}
v=-\dfrac bc-\frac{ca}{c+2}q\quad\mbox{if}\quad c\ne-2,0,\\
v=-\dfrac b2\ln|1-q^2|\quad\mbox{if}\quad c=0,\\
v=\dfrac b2+\frac a3q\ln|1-q^2|\quad\mbox{if}\quad c=-2.
\end{gather*}
In order to find solutions of~$\mathcal L$, we use the separation ansatz $v=e^{imp}\chi(q)$, where $m$ is a complex constant and $\chi$ is a complex-valued function of~$q$.
This ansatz reduces~$\mathcal L$ to the ordinary differential equation
\[
(1-q^2)\chi_{qq}- 2q\chi_q -\left(c+\frac{m^2}{1-q^2}\right)\chi=0.
\]

The reduced equation is the Legendre equation whose general solution is expressed in terms of the associated Legendre functions $P^m_n$ and $Q^m_n$ of first and second kind,
where the parameter~$n$ is a root of the equation $n(n+1)=-c$, i.e.\ $\chi = AP^m_n(q)+ BQ^m_n(q)$ with $A$ and $B$ being arbitrary complex constants.
The values~$m$, $\chi$, $n$, $A$ and~$B$ can be complex due to the linearity of~$\mathcal L$ as the real part of any solution of~$\mathcal L$ is also a solution of~$\mathcal L$.
Combining solutions obtained for a fixed value of~$n$ and different values of~$m$, we obtain, e.g. for $c=-n(n+1)\ne-2,0$, the solution
\[
 v(p,q) = \sum_{j=1}^N\mathop{\rm Re}\Big(A_je^{im_jp}P_n^{m_j}(q) + B_je^{im_jp}Q_n^{m_j}(q)\Big) - \frac{acq}{c+2}.
\]
(If $c\ne0$, the constant~$b$ can be neglected up to shifts of~$v$, which are Lie symmetry transformations for Eq.~\eqref{eq:RossbyHaurwitzClass} induced by shifts of~$\psi$.)
This gives rise to group-invariant solutions of the form
\[
 \psi(t,\lambda,\mu) = \sum_{j=1}^N\mathop{\rm Re}\Big(A_jP_n^{m_j}(\mu)e^{im_j(\lambda-(a-\Omega)t)}+B_jQ_n^{m_j}(\mu)e^{im_j(\lambda-(a-\Omega)t)}\Big) - \frac{ac\mu}{c+2}+\Omega\mu,
\]
where we already take into account the effects of rotation by reverting to the moving coordinate frame using transformation~\eqref{eq:PlatzmanTransformation}.

For the above solutions to be globally defined on the sphere, the parameters involved should satisfy special conditions.
Namely, we have that $n\in\mathbb N$, all $m_j$ take only integer values from~$-n$ to~$n$ and the constants $B_j$ are zero.
In other words, the globally defined solutions of the kind considered are expressed in terms of spherical harmonics of the same degree~$n$,
\begin{equation}\label{eq:GeneralizedRossbyHaurwitzWave}
 \psi(t,\lambda,\mu) = \sum_{m=0}^n A_{m}\cos(m\omega+\delta_m)P_n^m(\mu) - \frac{an(n+1)\mu}{n(n+1)-2}+\Omega\mu,
\end{equation}
where $n\in\mathbb N$ and $\omega=\lambda-(a-\Omega)t$, $A_m$ and $\delta_m$ are real constants,
$P_n^m$ is an associated Legendre polynomial, $P_n^m(\mu)=(-1)^m(1-\mu^2)^{m/2}(d^m/d\mu^m)P_n(\mu)$ and
$P_n$ is the ordinary Legendre polynomial of degree~$n$.
The solutions of this form can be interpreted as traveling waves in East--West direction superimposed to a West--East mean flow that only depends on the latitude. From the physical point of view, this class of solutions is reasonable as it neither violates the cyclic boundaries on the sphere nor has singular points (as would be the case for non-integer values of~$m$ and~$n$). Moreover, it describes a recurrent configuration for a barotropic atmosphere at mid-latitudes and it can be found qualitatively on almost every weather chart for the height of the 500 Hectopascal pressure surface.

\begin{figure}[htdp]
  \centering
  \includegraphics[scale=0.8]{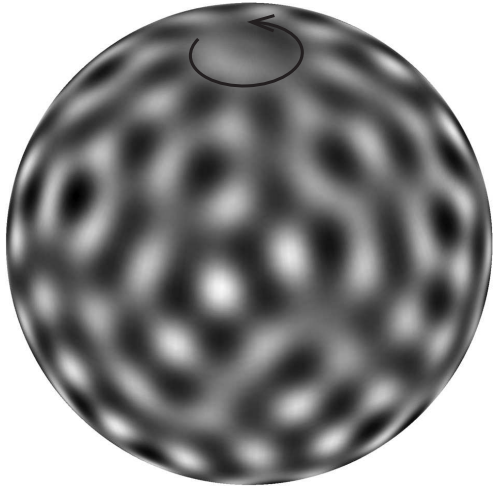}
  \caption{Plot of the generalized Rossby--Haurwitz wave solutions~\eqref{eq:GeneralizedRossbyHaurwitzWave} for $n=20$ at $t=0$, where $A_3=100$, $\delta_3=0$, $A_8=150$, $\delta_8=1.5$, $A_{13}=200$, $\delta_{13}=3.4$ and $A_{18}=250$, $\delta_{18}=0.9$. The remaining $A_{i}$, $\delta_{i}$ are identically zero. The solution evolves against the direction of rotation of the Earth in westward direction.}
  \label{fig:RossbyWave}
\end{figure}

In the more specific case when we require the mean flow to vanish, the relation $a = \frac{n(n+1)-2}{n(n+1)}\Omega$ must hold for the constants $a$ and $n$. In this case, the phase velocity of the waves is given by
\[
 c_{\mathrm{phase}} = -\frac{2\Omega}{n(n+1)},
\]
which is the well-known phase velocity of Rossby--Haurwitz waves. This is probably the most famous solution of the barotropic vorticity equation on the sphere. 
Evaluating the formula~\eqref{eq:GeneralizedRossbyHaurwitzWave} for a few nonzero values of $A_m$'s and a lower value of~$n$ already gives a quite nontrivial and physically reasonable form of the stream function~$\psi$, cf.\ Fig.~\ref{fig:RossbyWave}.

The reduction associated with the subalgebra $\langle\JJJ_1 + \ZZ(g) \rangle$ is of particular interest for the present paper letter. The reduction ansatz is $\psi = v(p,q) + g(t)\lambda$, where $p = t$, and $q = \mu$. The reduced form of Eq.~\eqref{eq:VorticityEquationSphere} is $w_p + gw_q = 0$, $w := ((1-q^2)v_q)_q$. This equation is completely integrable by quadratures, which leads to the family of exact solutions
\begin{equation}\label{eq:ExactSolutionRotationGauging}
 \psi=g(t)\lambda+f(t)+h(t)\mathrm{arctanh}\, \mu + \int\frac{\int w(\gamma)\ddd\gamma}{1-\mu^2}\,\ddd\mu,\qquad \gamma:=\mu-\int g(t)\ddd t
\end{equation}
of Eq.~\eqref{eq:VorticityEquationSphere}, where $f(t)$, $h(t)$ and $w(\gamma)$ are arbitrary functions of their arguments.

Let us now briefly discuss two-dimensional Lie reductions of Eq.~\eqref{eq:VorticityEquationSphere}.
An optimal list of two-dimensional subalgebras~$\mathcal S^\infty_0$ is
\begin{gather*}
    \langle \DDD + b\JJJ_1, \p_t \rangle, \qquad \langle \DDD, \JJJ_1 + \ZZ(at^{-1})\rangle, \qquad \langle \DDD + a\JJJ_1, \ZZ(|t|^b)\rangle, \\
    \langle\p_t, \JJJ_1 + \ZZ(c) \rangle, \qquad \langle\p_t + c\JJJ_1, \ZZ(e^{\tilde ct}) \rangle, \qquad
    \langle\JJJ_1 + \ZZ(\check g), \ZZ(\hat g) \rangle, \qquad  \langle\ZZ(g^1), \ZZ(g^2) \rangle,
\end{gather*}
where $a,b\in\mathbb R$, $c\in \{-1,0,1\}$; $\tilde c\in \{-1,0,1\}$ if $c=0$;
$\hat g$, $\check g$, $g^1$ and~$g^2$ are arbitrary smooth functions of~$t$ such that $g^1$ and~$g^2$ are linearly independent and $\hat g$ does not vanish.
We first note that Lie reductions using the third, the fifth, the sixth and the seventh algebra are not possible as they include subalgebras of the algebra $\langle \ZZ(g) \rangle$, which does not allow setting up a reduction ansatz. Then, the second and the fourth algebra include the subalgebra of the form $\langle\JJJ_1 + \ZZ(g) \rangle$ with the values $g=at^{-1}$ and $g=c$, respectively. As we have seen above, the reduced equation stemming from the algebra $\langle\JJJ_1 + \ZZ(g) \rangle$ can be integrated completely by quadratures for arbitrary~$g$. Therefore, no further Lie reduction with respect to two-dimensional subalgebras containing $\langle\JJJ_1 + \ZZ(g) \rangle$ must be carried out. It thus remains to investigate the Lie reduction corresponding to the first subalgebra $\langle \DDD + a\JJJ_1, \p_t \rangle$. An ansatz $\psi=e^{-\lambda/b}v(\mu)$ constructed using this subalgebra coincides, up to notation, with a particular case of the ansatz which is derived in the course of the reduction with respect to the subalgebra $\langle \p_t+a\JJJ_1\rangle$ (with $a=0$) and subsequent separation of variables in Eq.~\eqref{eq:RossbyHaurwitzClass} with linear~$F$.
Summing up, \emph{two-dimensional Lie reductions of Eq.~\eqref{eq:VorticityEquationSphere} give no new closed-form solutions in comparison with one-dimensional reductions. }

We also try to apply other tools for finding exact solutions to Eq.~\eqref{eq:VorticityEquationSphere}, such as partially invariant reductions and hidden symmetries.

In fact, the algorithm of partially invariant reduction cannot be directly used for Eq.~\eqref{eq:VorticityEquationSphere} since it is a single equation in a single unknown function, which is the stream function.
At the same time, usually we can replace a single differential equation by an equivalent system of differential equations with greater number of unknown functions.
There are at least two quite natural counterparts of the vorticity equation among systems of differential equations, namely, the corresponding Euler equations~\eqref{eq:NonlinearEulerEquationsSphere} for the flow velocity and the pressure and the vorticity equation itself,
where the vorticity is assumed as one more unknown function and its definition is treated as an equation relating~$\zeta$ and~$\psi$.
We consider the last system, which we will simply call system~\eqref{eq:VorticityEquationSphere}.
It is convenient that the maximal Lie invariance algebra~$\mathcal S^\infty_{0\rm s}$ of system~\eqref{eq:VorticityEquationSphere} is isomorphic to the algebra~$\mathcal S^\infty_0$.
More precisely, every vector field from $\mathcal S^\infty_{0\rm s}$ is a prolongation of an operator from $\mathcal S^\infty_0$ to the vorticity~$\zeta$ in view of its expression via the stream function~$\psi$.
This is why the lists of subalgebras presented in Section~\ref{sec:Subalgebras} can be used for finding partially invariant solutions of system~\eqref{eq:VorticityEquationSphere}.
It suffices to consider only subalgebras whose intersections with the ideal $\langle\ZZ(g)\rangle$ are equal to $\langle\ZZ(1)\rangle$.
We have tested two partially invariant reductions, with respect to the subalgebras $\langle\JJJ_1,\ZZ(1)\rangle$ and $\langle\DDD+a\JJJ_1,\p_t,\ZZ(1)\rangle$.
After cumbersome computation we have derived that all the corresponding partially invariant solutions are in fact Lie invariant.
Nevertheless, we plan to continue the study of partially invariant solutions for systems related to Eq.~\eqref{eq:VorticityEquationSphere} since this may possibly result in interesting solutions for Eq.~\eqref{eq:VorticityEquationSphere}.

Hidden symmetries are defined as Lie symmetries for submodels~\cite{abra06Ay}, and only purely hidden symmetries, which are not induces by Lie symmetries of the initial model, are interesting.
Unfortunately, all Lie symmetries of both the valuable reduced equations~\eqref{eq:VorticityEquationSphereFirstReduction} and~\eqref{eq:RossbyHaurwitzClass} are induced by symmetries of Eq.~\eqref{eq:VorticityEquationSphere}.
At the same time, the homogeneous linear counterpart~$\mathcal L$ for the once integrated equation~\eqref{eq:RossbyHaurwitzClass} admits, in addition to the induced Lie symmetry $\p_p$,
the Lie symmetries $v\p_v$ and $h(p,q)\p_v$ which arise due to the linearity of~$\mathcal L$ and are purely hidden for Eq.~\eqref{eq:VorticityEquationSphere}.
Here the function $h=h(p,q)$ traverses the set of solutions of the equation~$\mathcal L$.
The above symmetry allows us to separate variables and to linearly combine solutions.
This is the fact which explains from the symmetry point of view why Lie reduction with respect to the class of subalgebras $\langle \p_t+a\JJJ_1\rangle$ gives the widest and most interesting family of solutions.

Let us now compare the reductions from~\cite{bihl09Ay} discussed above with those carried out in~\cite{ibra11Ay}, where Lie reductions were computed for two particular cases of two-dimensional subalgebras. It turns out that these cases are equivalent to each other and, moreover, can be derived as special instances of the exact solution~\eqref{eq:ExactSolutionRotationGauging} constructed in the course of Lie reduction with respect to the one-dimensional subalgebra $\langle\JJJ_1 + \ZZ(g) \rangle$.

Explicitly, the first case is the Lie reduction with respect to the algebra similar to $\langle\DDD,\JJJ_1\rangle$.
The corresponding solution
\[
 \hat \psi^{[1]} = \frac{\cos\theta}{2\nu R_0}-\frac{H(\theta)}{\nu} + \frac{C_1}{t}+\frac{C_2}{t}\ln\left|\tan\frac\theta2\right|,
\]
where $H'(\theta)=F(\theta)$, is included as a particular instance in the exact solution~\eqref{eq:ExactSolutionRotationGauging} by setting
\[
 g(t)=0,\quad f(t)=\frac{C_1}{\nu t},\quad h(t)=-\frac{C_2}{\nu t},\quad w(\gamma)=0.
\]
Note that the first term in the solution $\hat \psi^{[1]}$ is not essential as it accounts for the rotation of the reference frame which can be taken into account for the solution~\eqref{eq:ExactSolutionRotationGauging} in the nonrotating reference frame upon applying the transformation~\eqref{eq:PlatzmanTransformation}. The second term in $\hat \psi^{[1]}$ arises due to invoking the stream function only for the perturbation velocities and can be removed upon using the transformation~\eqref{eq:RelationStreamFunctions}.

The second closed-form solution was constructed in~\cite{ibra11Ay} in the course of reduction using the algebra similar to~$\langle\DD,\JJJ_3\rangle$ and it reads
\[
 \hat \psi^{[2]}=\frac{\cos\theta}{2\nu R_0}-\frac{H(\theta)}{\nu} + \frac{C_1}{t}+\frac{C_2}{t}\ln\Big|\frac{1-\kappa}{1+\kappa}\Big|,
\]
where $\kappa=\sin\theta\cos(\lambda+t/(2R_0))$. However, this solution is related to the above solution~$\hat \psi^{[1]}$ and hence to the solution~\eqref{eq:ExactSolutionRotationGauging} via a point transformation. Namely, from the adjoint action
\[
 \Ad(e^{\ve\JJJ_2})\JJJ_1=-\JJJ_3\sin\ve+\JJJ_1\cos\ve,
\]
it is clear that for the value $\ve=-\pi/2$, the operator $\JJJ_1$ is mapped to $\JJJ_3$. The same transformation then maps the solution~\eqref{eq:ExactSolutionRotationGauging} to the solution including the particular form~$\hat \psi^{[2]}$. Explicitly, the finite transformation component of $\JJJ_2$ for $\mu$ is $\tilde \mu=\mu\cos\ve+\sqrt{1-\mu^2}\cos\lambda\sin\ve$, which for $\ve=-\pi/2$ gives $\tilde \mu=-\sqrt{1-\mu^2}\cos\lambda$. The transformed form of~\eqref{eq:ExactSolutionRotationGauging} for $g(t)=0$ then takes the form
\[
 \psi=f(t)-h(t) \mathrm{arctanh}\,(\sqrt{1-\mu^2}\cos\lambda) + W(\sqrt{1-\mu^2}\cos\lambda),
\]
where $W=W(\sqrt{1-\mu^2}\cos\lambda)$ is an arbitrary function of its argument. The solution~$\hat \psi^{[2]}$ is the particular case of the above solution where
\[
 f(t)=\frac{C_1}{\nu t},\quad h(t)=-\frac{C_2}{\nu t},\quad W=0.
\]
Again, the first two terms in~$\hat \psi^{[2]}$ are inessential in that they can be added upon transforming to the rotating coordinate system and redefining the stream function.

In short, the two exact solutions $\hat \psi^{[1]}$ and $\hat \psi^{[2]}$ and all the special forms therefrom studied in~\cite{ibra11Ay} can be united upon investigating the solution~\eqref{eq:ExactSolutionRotationGauging} stemming from the algebra~$\langle\JJJ_1+\ZZ(g)\rangle$, which belongs to the list of inequivalent one-dimensional subalgebras of the maximal Lie invariance algebra~\eqref{eq:MaximalLieInvarianceAlgebraVorticityEquationSphere}.

\section{Concluding remarks}\label{sec:Conclusion}

The present letter is devoted to the relation of the two independent studies~\cite{bihl09Ay} and~\cite{ibra11Ay} on a particular form of the two-dimensional incompressible nonlinear Euler equations in a rotating reference frame in spherical geometry. The main intention of this work is to point out the importance of systematically following the algorithm of Lie reduction in order to arrive at inequivalent exact solutions of a system of differential equations. The key aspects of this algorithms are the following:
\begin{enumerate}\itemsep=0ex
\renewcommand{\labelenumi}{(\roman{enumi})}

 \item Identify essential and inessential parameters in the system of differential equations at hand. This point is systematically addressed by solving the group classification problem, see e.g.~\cite{ovsi82Ay,popo10Ay}. In the case of the vorticity equation in a rotating reference frame~\eqref{eq:VorticityEquationSphere} the solution of the group classification problem gives that $\Omega$ can be set to zero upon using the simple point transformation~\eqref{eq:PlatzmanTransformation}.

 \item Determine an optimal list of inequivalent subalgebras of the maximal Lie invariance algebra of the system of differential equations of interest which are suitable for Lie reduction. This task usually requires to classify low-dimensional inequivalent subalgebras under the adjoint action of a Lie group on its Lie algebra, see e.g.~\cite{andr98Ay,olve86Ay,ovsi82Ay} for the relevant techniques involved in this step. It is of utmost importance to do this classification, as the result will give information on what Lie reductions should be carried out for the particular system of differential equations. In the case of the vorticity equation~\eqref{eq:VorticityEquationSphere} a suitable optimal system was found in~\cite{bihl09Ay}. As no such optimal system was established in~\cite{ibra11Ay}, the Lie reductions that were carried out are not guaranteed to give inequivalent solutions. Indeed, as is shown in Section~\ref{sec:RelationAlgebrasIbragimovBarotropicVorticity} the two-dimensional reductions studied in~\cite{ibra11Ay} are equivalent to each other with respect to Lie symmetry transformations of Eq.~\eqref{eq:VorticityEquationSphere}.

 \item If applicable, study the physical importance of the solutions obtained. The question of physical meaning usually cannot be answered solely in the framework of group analysis. However, what symmetries can add to this investigation is that any solution obtained in the course of Lie reduction is in fact only a representative of a parameterized family of similar solutions, which can be obtained upon acting with symmetry transformations on the derived solution. In some cases, an unphysical exact solution is simply the consequence of choosing an inappropriate ansatz for Lie reduction.
     \looseness=-1

\end{enumerate}

Depending on the complexity of the maximal Lie invariance algebra of the system of differential equations under consideration, the exhaustive solution of the Lie reduction problem can be a tedious task. It might therefore be tempting to compute only the Lie symmetries of a differential equation (which is an algorithmic task that can be handled with the aid of computer algebra) and then to use certain obvious combinations of the infinitesimal generators in order to carry out the reduction. Combinations that arise quite often for equations in sciences are those leading to scale-invariant or wave solutions. On the other hand, even if such solutions might have an immediate physical interpretation, there is no guarantee that they are not equivalent to each other. A systematic approach to Lie reduction therefore cannot avoid the classification of inequivalent subalgebras.

Another issue that arises in the course of Lie reduction and which can considerably simplify the solution of the Lie reduction problem is whether one is able to completely integrate the equations arising at a low stage of reduction. In this case, the reduction with respect to higher-dimensional algebras that include as a subalgebra the respective algebra which enabled the integration can be omitted. This holds true in the case of the two-dimensional Lie reductions carried out in~\cite{ibra11Ay}, which all follow from a one-dimensional reduction of the vorticity equation that can already be integrated completely.

Lie reduction is only one method to derive exact solutions for systems of differential equations. The notion of point symmetries can be extended beyond point character and point transformations can be used in more sophisticated methods. Other techniques that can lead to valuable physical solutions are e.g.\ conditional symmetries, potential symmetries, nonclassical symmetries or the method of differential constraints. In order to properly apply these techniques, however, an exhaustive understanding of the possible Lie reductions is of certain value. This holds in particular for the techniques that are extensions of the classical Lie method, such as the method of differentially invariant solutions, the direct (ansatz) method, which is equivalent to the reduction method using conditional symmetries, or partially invariant reductions with Lie symmetries. Note that the latter have been tested for the vorticity equation on the sphere in this letter and it was found that they reduce to the group-invariant solutions that we found above.

From the physical point of view, deriving exact solutions of atmospheric models in spherical geometry is a task of immediate relevance as such solutions can serve as important benchmark tests for numerical general circulation models. As an example for such benchmarks, in~\cite{will92Ay} a suite of seven tests was proposed for the assessment of numerical models for the shallow-water equations in spherical coordinates. By now, virtually every new numerical model of the shallow-water equations is tested using the suggested single benchmarks, all of which allow the evaluation of different relevant aspects of the shallow-water equations. By providing a host of exact and physically relevant solutions for the barotropic vorticity equation in spherical geometry, with this letter we aim to extend the list of known solutions of models of geophysical fluid dynamics on the rotating sphere. We should like to stress in particular that using the method of Lie reduction, several exact solutions can be derived that extend or generalize known solutions of the vorticity equation, such as the well-known Rossby--Haurwitz wave solution. It is therefore hoped that the general classes of exact solutions found will be of further relevance in the field of numerical geophysical fluid dynamics.

\section*{Acknowledgements}

This research was supported by the Austrian Science Fund (FWF), projects J3182--N13 (AB) and P20632 (ROP).

\end{document}